\documentclass[conference]{IEEEtran}
\IEEEoverridecommandlockouts
\usepackage{cite}
\usepackage{amsmath,amssymb,amsfonts}
\usepackage{algorithmic}
\usepackage{color,url,graphicx}
\usepackage{multirow}
\usepackage{textcomp}

\makeatother
\begin{document}
\title{Robust Image Identification for Double-Compressed JPEG Images}

\IEEEpubid{ \makebox[\columnwidth]{978-1-5386-2350-3/18/\$31.00 {\copyright} 2018 IEEE\hfill} \hspace{\columnsep}\makebox[\columnwidth]{} }\author{\IEEEauthorblockN{Kenta Iida\textsuperscript{1} and Hitoshi Kiya\textsuperscript{1}}
\IEEEauthorblockA{\textsuperscript{1}Tokyo Metropolitan University, Tokyo, Japan}
\IEEEauthorblockA{Contact author e-mail: kiya@tmu.ac.jp } 
}
\maketitle
\begin{abstract}
It is known that JPEG images uploaded to social networks (SNs) are mostly re-compressed by the social network providers.
Because of such a situation, a new image identification scheme for double-compressed JPEG images is proposed in this paper.
The aim is to detect single-compressed images that have the same original image as that of a double-compressed one.
In the proposed scheme, the signs of only DC coefficients in DCT coefficients and one threshold value are used for the identification.
The use of them allows us to robustly avoid errors caused by double-compression, which are not considered
in conventional schemes.
The proposed scheme has applications not only to find uploaded images corresponding to double-compressed ones, but also to detect some image integrity.
The simulation results demonstrate that the proposed one outperforms conventional ones including state-of-art image hashing one in terms of the querying performance.
\end{abstract}


\IEEEpeerreviewmaketitle

\section{Introduction}
\label{sec:intro}
The growing popularity of social networks (SNs) like Twitter and Facebook has opened new perspectives in many research fields, including the emerging area of multimedia forensics.
The huge amount of images uploaded to SNs are generally stored in a compressed format as JPEG images, after being re-compressed using compression parameters different from those used for the uploaded images\cite{Intro,Intro1,Intro2}.
Due to a such situation, identifying JPEG images which have the same original image is required.

Several identification schemes and hash functions have been proposed \cite{id1,id2,id4,conv,conv2,fcs1,fcs2,zpid,zpid2,cbir,ih1,ih2,itq}.
They have been developed for the various purposes:producing evidence regarding image integrity, robust image retrieval, finding illegally distributed images and so on.
However, those schemes do not assume the double-compressed images.
Therefore, for the applications using images uploaded to SNs, a new scheme for double-compressed images is required.

The conventional schemes for identifying images can be broadly classified into two types: compression-method-dependent and compression-method-independent. 
Compress-ion-method-independent schemes include image retrieval and image hashing-based ones \cite{cbir,ih1,ih2,itq}.
These schemes generally extract features (e.g. GIST\cite{gist} and SURF\cite{surf}) after decoding images,  and then the features are converted to other representations.
In some schemes, several noises including errors caused by lossy compression are considered.
However, they sometimes miss slightly differences because they mainly aim to retrieve similar images.

On the other hand, due to the use of robust features against JPEG errors, compression-method-dependent schemes \cite{conv,conv2,fcs1,fcs2,zpid,zpid2} generally have stronger robustness against JPEG errors than first type ones.
The schemes\cite{conv, fcs1} use positive and negative signs of discrete cosine transform (DCT) coefficients, and the scheme\cite{zpid,zpid2} uses the positions in which DCT coefficients have zero values and quantization matrices.
For the identification between single-compressed images, they have a high querying performance even under the various conditions, and it is guaranteed that there are any false negative matches in principle.
However,  they do not consider JPEG errors in double-compressed images.  
  
Due to such situations, our proposed scheme can robustly identify JPEG images double-compressed under various compression conditions.
The proposed scheme uses the signs of only DC coefficients as a feature.
In addition, considering the DC sign inversion sometimes caused by double-compression, the use of one threshold value is considered. 
This strategy allows us to identify images with avoiding errors caused by double-compression. 
In the simulations, the proposed scheme is compared with several  conventional ones including the state-of-art image hashing scheme\cite{ih1,itq,zpid} in terms of querying performances.
The results demonstrate that the proposed scheme enables to detect slightly differences and outperforms conventional schemes, even if images are very similar.

\section{Preliminaries}
\subsection{JPEG Encoding}
The JPEG standard is the most widely used image compression standard.
The JPEG encoding procedure can be summarized as follows.

\begin{itemize}
  \item[1)]
Perform color transformation from RGB space to $\mathrm{YC_{b}C_{r}}$ space and sub-sample $\mathrm{C_{b}}$ and $\mathrm{C_{r}}$.
 \item[2)]
Divide an image into non-overlapping consecutive 8$\times$8-blocks.

 \item[3)]
Apply DCT to each block to obtain 8$\times$8 DCT coefficients $\bf S$, after mapping all pixel values in each block  from [0,255] to [-128,127] by subtracting 128 in general.

 \item[4)]
Quantize $\bf S$ using a quantization matrix $\bf Q$.

 \item[5)]
Entropy code using Huffman coding.
\end{itemize}
After applying DCT to the shifted values in each block, 64 DCT coefficients ${\bf S}$ in each block, namely one DC and 63 AC coefficients, are obtained.
The range of the DC coefficient is [-1024,1016].

In step 4), a quantization matrix ${\bf Q}$ with 8$\times$8 components is used to obtain a matrix $\bf S_{q}$ from $\bf S$.
For example,
\begin{equation}
\label{eq:quant}
S_{q} (u, v)=\mathrm{round}\left(\frac{S(u, v)}{Q(u, v)}\right),\ 0\leq u\leq 7,\  0\leq v \leq 7,
\end{equation}
where $S(u, v)$, $Q(u, v)$ and $S_{q}(u, v)$ represent the $(u,v)$ element of $\bf S$, $\bf Q$ and $\bf S_{q}$ respectively.
The $\mathrm{round}(x)$ function is used to round a value $x$ to the nearest integer value and $\lfloor x \rfloor$ denotes the integer part of $x$. 

The quality factor $QF\ (1\leq QF \leq 100)$ parameter is used to control a matrix $\bf Q$.
The large $QF$ results in a high quality image.

\subsection{Image Manipulation on Social Networks\label{sec:sn}}
Let consider JPEG images are uploaded to SNs.
It is known that JPEG images uploaded to SNs are often manipulated as below\cite{Intro,Intro1,Intro2}.
\begin{itemize}
 \item[a.]Editing metadata and the filename\\
Most of metadata in the header are deleted for privacy-concerns and the filenames of uploaded images are changed.
 \item[b.]Re-compressing uploaded images by JPEG\\
Before stored in a database of SNs, uploaded images are decoded once and then the images are compressed again under the different coding condition.
 \item[c.]Resizing uploaded images\\
If uploaded images satisfy certain conditions, those images are resized.
For instance, in Twitter, when the filesize of images is larger than 3MB or the size of images is larger than 4096$\times$4096, the images will be resized.
\end{itemize}
Due to these manipulations, robust content-based methods against the errors caused by re-compression are required to find the uploaded images corresponding to downloaded one.
In this paper, it is assumed that the size of uploaded images is the same as that of downloaded ones.

\subsection{Scenario}
\label{sec:id}
Let us consider a situation in which there are two or more compressed images generated under different or the same coding conditions.
They originated from the same image and were compressed under the various coding conditions.
We refer to the identification of those images as ``image identification".
Note that the aim of the image identification is not to retrieve visually similar images.

\begin{figure}[t]
\begin{center}
\begin{tabular}{c}
\begin{minipage}{\hsize}
  \begin{center}
   \includegraphics[width=85mm]{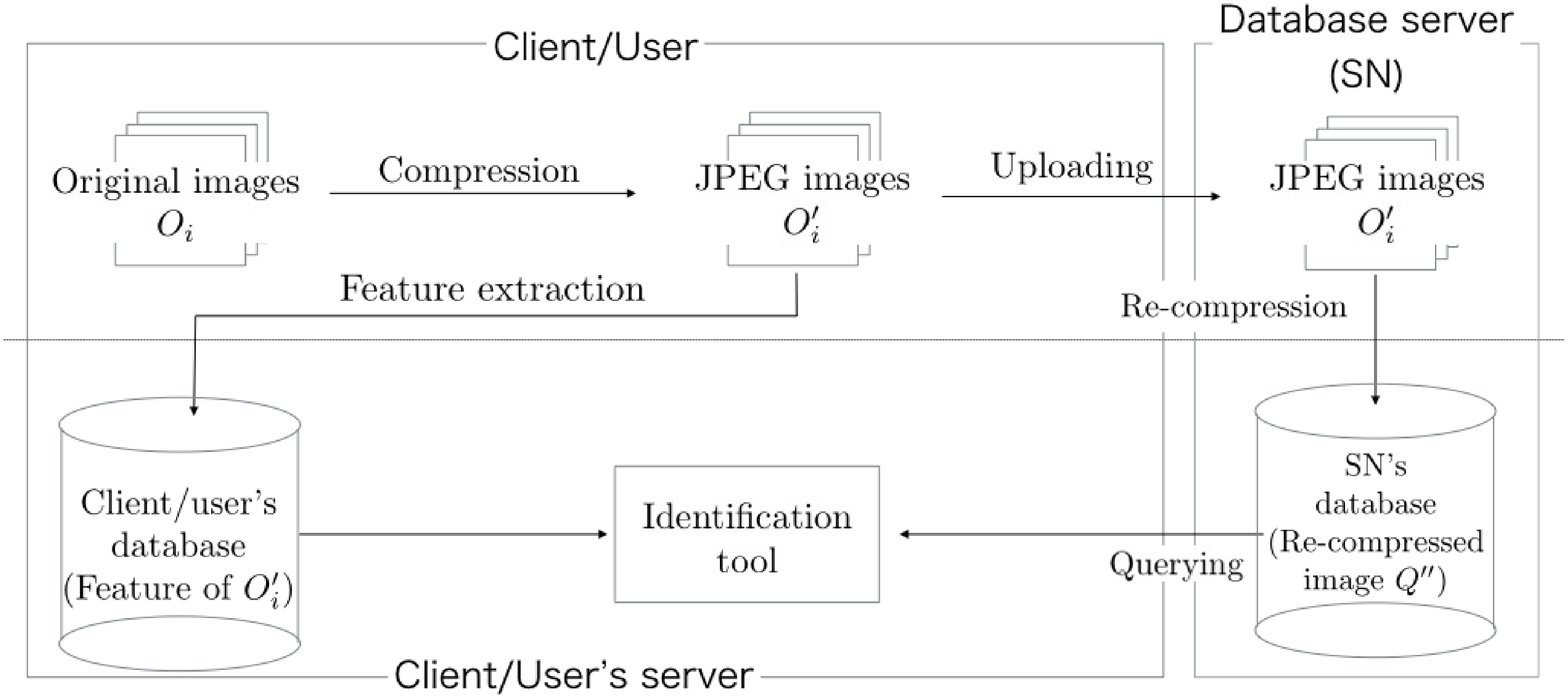}
      \end{center}
 \end{minipage}\\
\end{tabular}
\caption{Scenario}
 \label{fig:system}
 \end{center}
\end{figure}

The scenario of this paper is illustrated in Fig. \ref{fig:system}.
In this scenario,  a client/user identifies images by using an identification tool.
When the client/user uploads JPEG images to a database server like Twitter, the features of these images are enrolled (extracted and then stored) in a client/user's database.
The uploaded images are re-compressed under different coding parameters and then are stored in a database server.
Finally, the client/user carries out the identification after extracting the feature from a query image i.e. an uploaded image.

\begin{figure}[t!] 
   \centering
   \includegraphics[width=8.5cm]{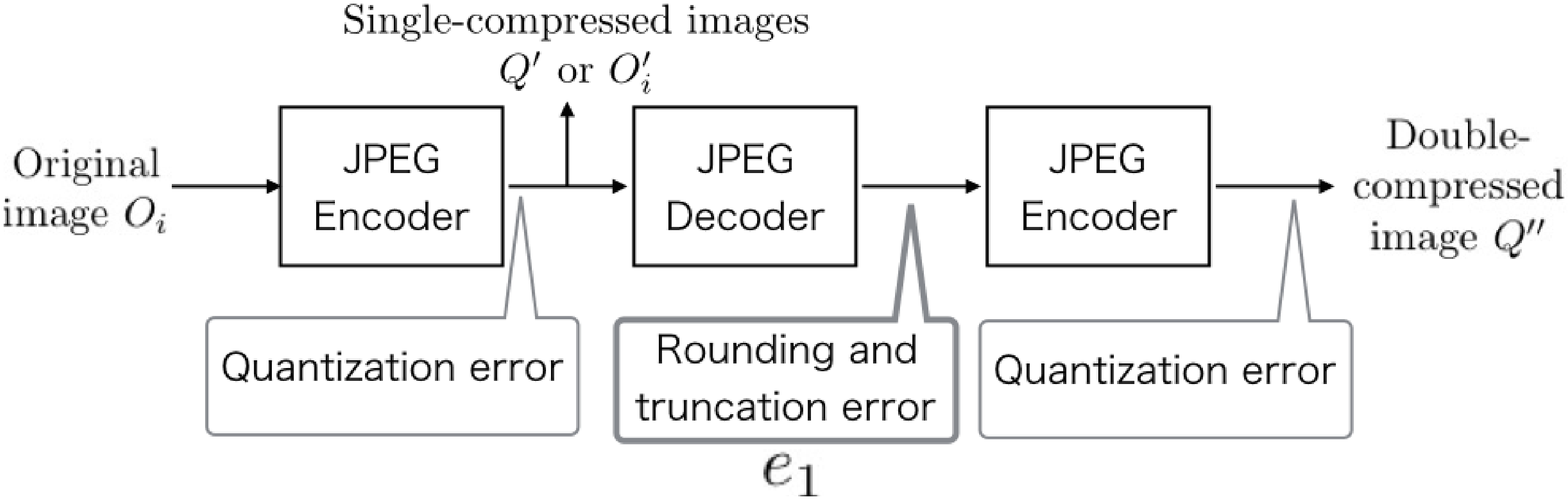} 
   \caption{JPEG errors in single-/double-compression}
   \label{fig:comp}
\end{figure}

The JPEG standard is generally used as a lossy compression method, so that several errors are caused in the generation process of double-compressed images\cite{dj}, as shown in Fig.\ref{fig:comp}.
In addition to ``quantization error" in the encoding process, ``rounding and truncation error" i.e. $e_1$ is caused in the decoding process.
In the proposed scheme, the errors in both processes are considered to identify double-compressed images.

\subsection{Notations and Terminologies}
The notations and terminologies used in the following sections are listed here.
\begin{itemize}
 \item
$O'_i$  represents a single-compressed image of an original image $O_i$.
 \item
$Q'$ represents a single-compressed query image and $Q''$ represents a double-compressed query image generated from $Q'$ (all images including $O_i'$ have the same size). 
 \item
$M$ represents the number of 8$\times$8-blocks in an image. 
 \item
$O'_i(m, n)$, $q'(m, n)$ and $q''(m, n)$ indicate $n$th quantized DCT coefficient in $m$th block in images $O'_i$, $Q'$ and $Q''$ respectively ($0 \leq m < M$, $0\leq n < 64$).
 \item
$QF_{O'_i}$, $QF_{Q'}$ and $QF_{Q''}$ indicate quality factors used to generate $O'_i,\ Q'$ and $\ Q''$ respectively. 
 \item
$\mathrm{sgn}(y)$ represents the sign of a real value $y$ as
\begin{equation}
\mathrm{sgn}(y)= \left \{
\begin{array}{c}
1,\  y>0,\\
0,\  y=0,\\
-1,\  y<0.\\
\end{array}
\right.
\end{equation}
\end{itemize}

\subsection{Property of DCT Coefficients}
It is verified from Eq.(\ref{eq:quant}) that quantized DCT coefficients have the following property\cite{conv,fcs1}.
\begin{itemize}
\item
When single-compressed images $Q'$ and $O'_{i}$ are generated from the same original image $O_i$, the positive and negative signs of DCT coefficients of the two images are equivalent in the corresponding location.
Namely, the relation is given as
\begin{equation}
\label{eq:sign}
\mathrm{sgn}(q'(m, n))= \mathrm{sgn}(O'_{i}(m, n)),
\end{equation}
where $q'(m, n)\neq 0$ and $O_i'(m, n)\neq 0$.
\end{itemize}
Assuming the identification of a double-compressed image i.e. $Q''$, this relation is also satisfied in all locations if $e_1=0$. However, $e_1 \neq 0$ in practical, so that the relation may be unsatisfied in a few locations.

In the proposed scheme, only the signs of DC coefficients are used as a feature of JPEG images.
Compared to the case of using all DCT coefficients, this feature can be compactly stored.
Besides, it is expected that DC signs have robustness against the rounding and truncation error, compared to the case of using those of AC coefficients.
This is because DC coefficients generally have the large absolute values, although a lot of AC coefficients do not.

\section{Proposed Scheme\label{sec:proposed}}
The proposed scheme uses the signs of DC coefficients in Y component and one threshold value for identification. 
The use of them allows to identify images double-compressed under various coding conditions.
Enrollment and identification processes are explained, here.

\subsubsection*{1) Enrollment Process}
\noindent In order to enroll the feature of image $O'_i$,  a client/user carries out the following steps.
\begin{itemize}
 \item[(a)]
Set values $M$ and $th$, where $th$ represents a threshold value used for the feature extraction.
 \item[(b)]
Set $m:=0$.
 \item[(c)]
Map a DCT coefficient $O_i'(m, 0)$ into $v_{O_i'}(m)$ as  
\begin{equation}
\label{eq:map}
v_{O'_i}(m)= \left \{
\begin{array}{l}
1,\  O'_i(m, 0) > th,\\
-1,\  O'_i(m, 0) < -th,\\
0,\  -th  \leq O'_i(m, 0) \leq th,\\
\end{array}
\right.
\end{equation}
where ${O'_i}(m,0)$  represents a DC coefficient in $m$th block of image $O'_i$.
 \item[(d)]
Set $m:=m+1$.
If $m < M$, return to step (c). 
Otherwise, store ${\bf v}_{O'_i}$ as the feature  in the client/user's database.
\end{itemize}

\subsubsection*{2) Identification Process}
In order to compare image $Q''$ with image $O'_i$, the client/user carries out the following steps.
\begin{itemize}
 \item[(a)]
Set value $M$.
 \item[(b)]
Set $m:=0$ . 
 \item[(c)]
If $v_{O'_i}(m)=0$ or $\mathrm{sgn}(q''(m,0))=0$, proceed to step(e).
 \item[(d)]
If $v_{O'_i}(m)\neq \mathrm{sgn}(q''(m,0))$, the client/user judges that $O'_i$ and $Q''$ are generated from different original images and the process for image $O'_i$ is halted.
 \item[(e)]
Set $m:=m+1$.
If $m < M$, return to step (c).
Otherwise, the client/user judges that $O'_i$ and $Q''$ are generated from the same original image.
\end{itemize}
In the proposed scheme, a threshold value $th$ is used to avoid the effect of double-compression i.e. $e_1$.

\section{Simulation}
A number of simulations were conducted to evaluate the performance of the proposed scheme.
\subsection{Selection of Threshold Value}
At first, to select the threshold value $th$, we conducted a preliminary experiment.
For this experiment, 885 images (the size of 384$\times$512) in Uncompressed Color Image Database (UCID)\cite{ucid} were compressed with $QF$=70, 75, 80, 85, 90, 95 respectively,  and then each image was re-compressed using $QF$=70, 75, 80, 85, 90, 95 respectively.
Comparing the DC coefficients of each single-compressed image with those of corresponding double-compressed ones, we determined the threshold value to avoid the effect of $e_1$.

It was observed from this experimental result that the signs were inverted when the absolute values of DC coefficients in a single-compressed image were smaller than 14. 
Therefore, $th=14$ was used for the identification.

\subsection{Querying Performance}
\begin{figure}[t!]
\begin{center}
\begin{tabular}{c}
 \begin{minipage}{0.3\hsize}
  \begin{center}
   \includegraphics[width=23mm]{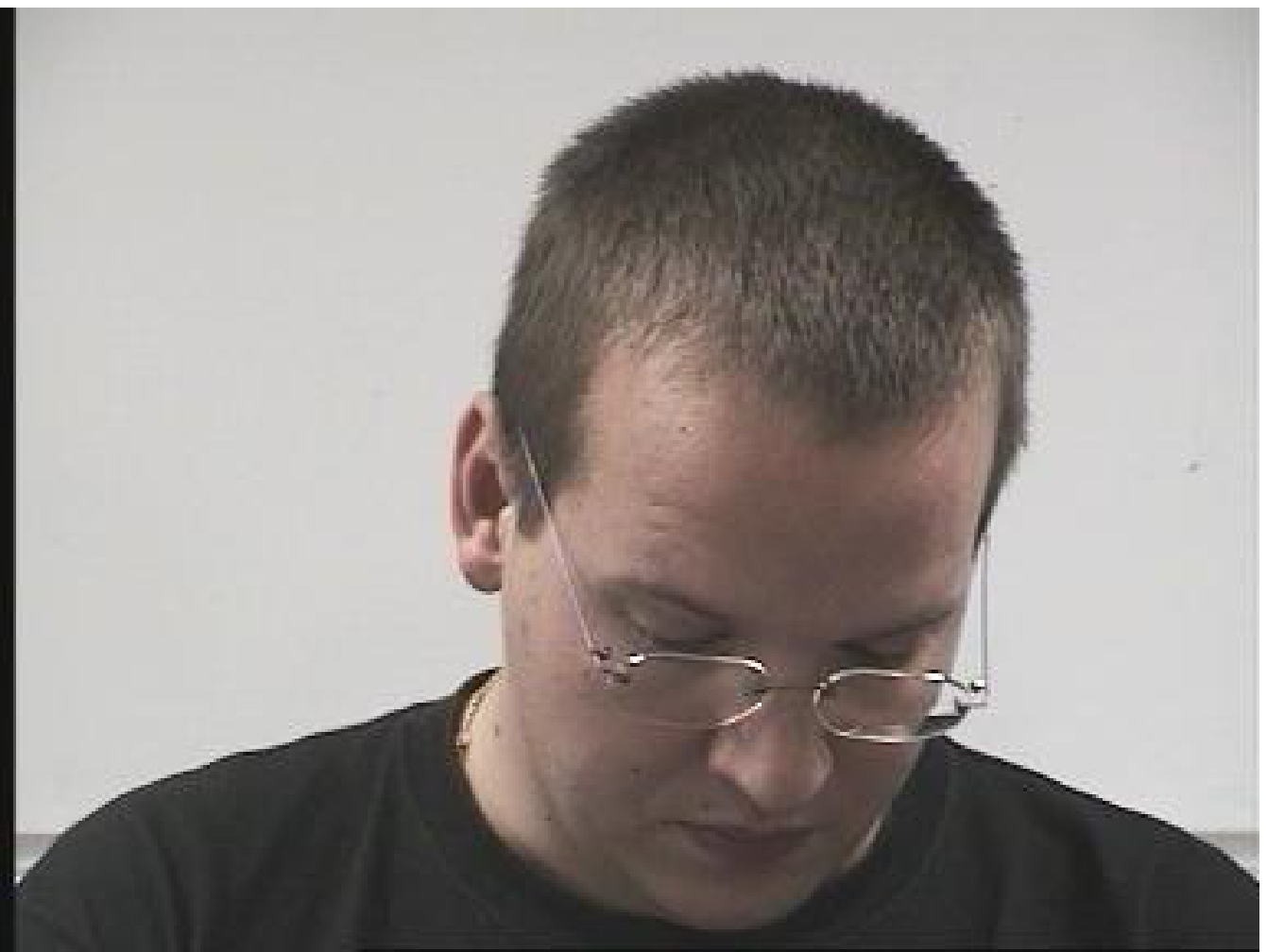}
  \end{center}
 \end{minipage}
  \begin{minipage}{0.3\hsize}
  \begin{center}
   \includegraphics[width=23mm]{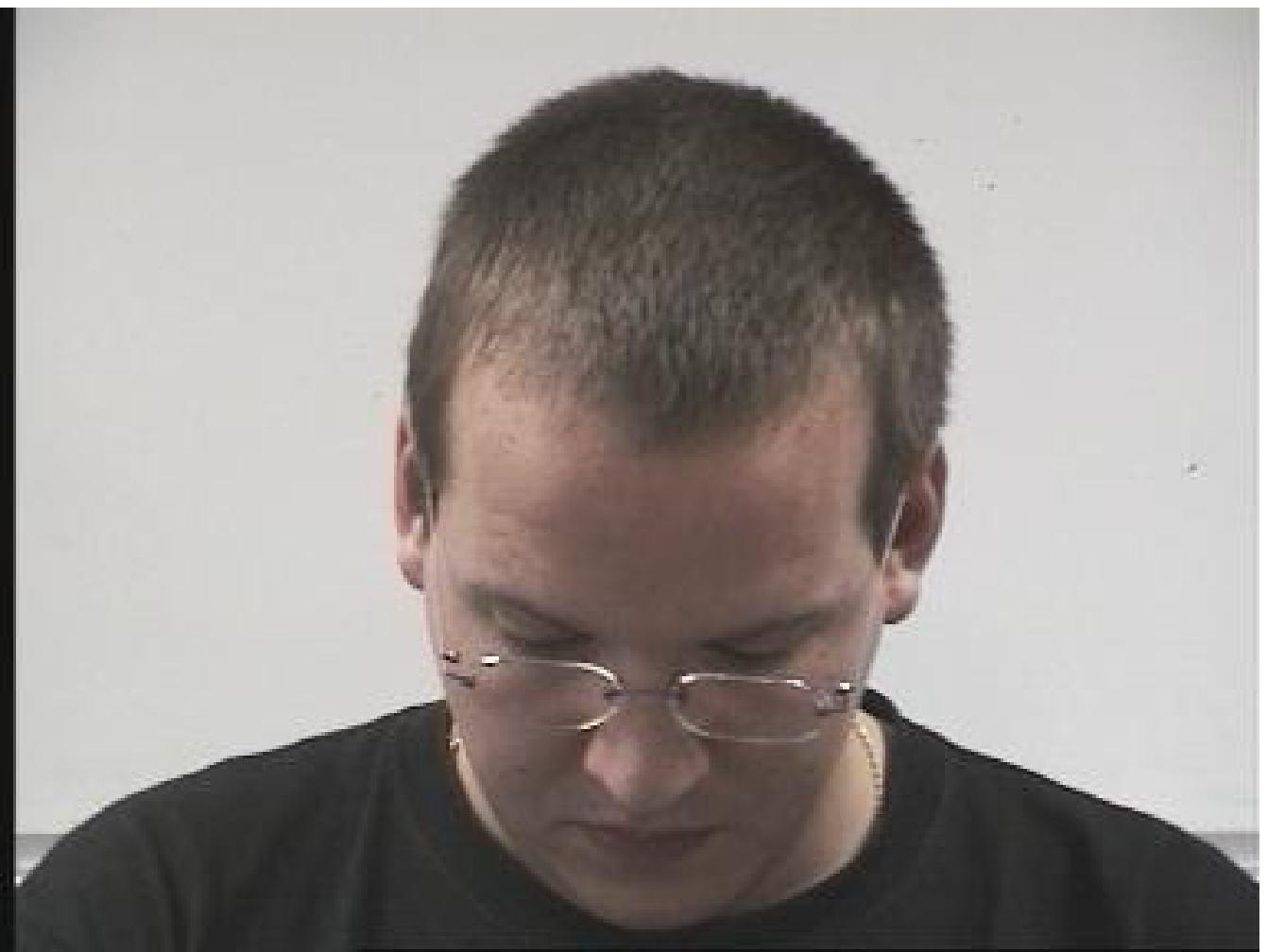}
  \end{center}
 \end{minipage}
  \begin{minipage}{0.3\hsize}
  \begin{center}
   \includegraphics[width=23mm]{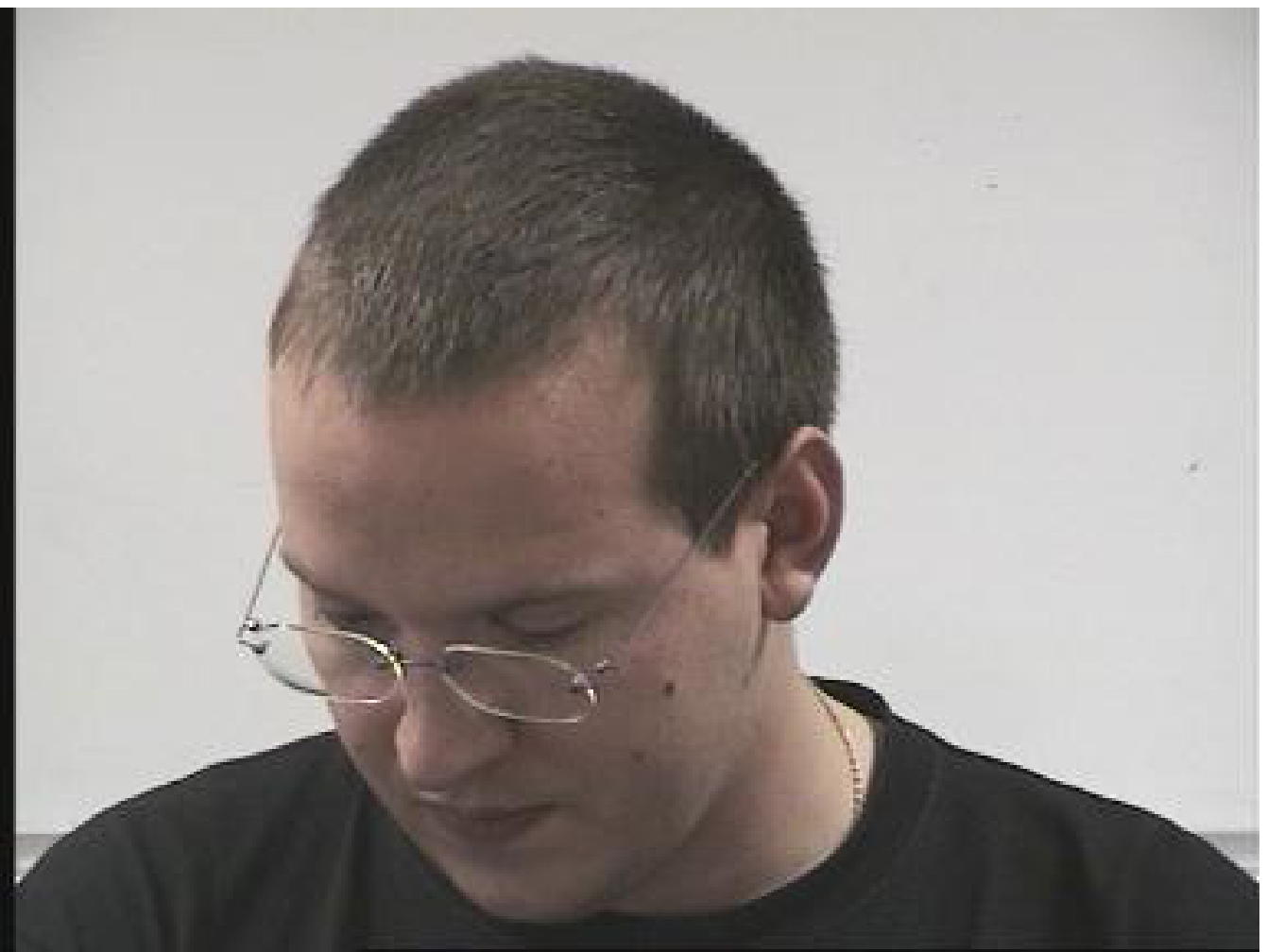}
  \end{center}
 \end{minipage}
 \end{tabular}
\caption{Examples of test images (288$\times$384)}
\label{fig:exhpid}
 \end{center}
\end{figure}

Next, we used the images in Head Pose Image Database (HPID) \cite{hpid} to evaluate querying performance between single-compressed images and double-compressed ones.
HPID consists of very similar images, as shown in Fig. \ref{fig:exhpid}. 
The main reason of using HPID is to show that the proposed scheme can detect a slight differences between the images.
Therefore, we used 186 images of ``Person01'' in HPID as original images. 

\begin{table}[t!]
\caption{Quality factors used to generate JPEG images\label{tab:condition}.$DB_1$, $DB_2$ and $DB_3$ indicate databases of client/user in Fig.\ref{fig:system}}
\centering
\scalebox{.9}{
\begin{tabular}{|c|c|c|c|c|c|}\hline
\multicolumn{2}{|c|}{JPEG images}&Quality factors\\\hline
\multirow{3}{*}{\shortstack{Images stored as \\features in}}&$DB_1$&$QF_{O'_i}=95$\\\cline{2-3}
&$DB_2$&$QF_{O'_i}=85$\\\cline{2-3}
&$DB_3$&$QF_{O'_i}=75$\\\hline
\multicolumn{2}{|c|}{Query images ($Q''$)} &$QF_{Q''}=71,75,80,85$($QF_{Q'}=QF_{O'_i}$)\\\hline
\end{tabular}
}\end{table}

Table \ref{tab:condition} summarizes the quality factors used to generate JPEG images, where $DB_1$, $DB_2$ and $DB_3$ indicate the databases of client/user in Fig.\ref{fig:system}.
For instance, features stored in the database $DB_1$ were extracted from 186 images compressed with $QF_{O'_i}=95$.
As query images for $DB_1$, 186 images with $QF_{O'_i}=95$ were re-compressed with $QF_{Q''}=71,75,80,85$ respectively.
It is known that the range of quality factors used for re-compression in SNs is [71,85] as in \cite{Intro2}, so that these quality factors were used.
As a result,  744 query images were used for $DB_1$.
Similarly, features in $DB_{2}$ and $DB_{3}$ were extracted from images with $QF_{O'_i}=85$ and $QF_{O'_i}=75$ respectively, and then images in each database were re-compressed with $QF_{Q''}=71,75,80,85$.  
Therefore, identification was performed 186$\times$744 times for each database to evaluate the proposed scheme.

The proposed scheme was compared with zero value positions-based scheme\cite{zpid} and two image hashing-based schemes: low-rank and sparse decomposition-based scheme\cite{ih1} and iterative quantization(ITQ)-based one\cite{itq}, where ITQ-based hash values were generated from 512 dimensional GIST feature vectors and each hash value was represented by 256 bits.  
In the schemes\cite{ih1,itq}, the hamming distances between the hash value of a query image and those of all images in each database were calculated, and then images that had the smallest distance were chosen as the images generated from the same original image as the query, after decompressing all images.

Table \ref{tab:res}  shows Precision $p$ and Recall $r$,  defined by
\begin{equation}
p = \frac{TP}{TP+FP},\ r = \frac{TP}{TP+FN},
\end{equation}
where TP, FP and FN represent the number of true positive, false positive and false negative matches respectively.
Note that $r=100[\%]$ means that there were no false negative matches and $p=100[\%]$ means that there were no false positive matches.

It is confirmed that only the proposed scheme did not provide any false and negative matches, although the other schemes did.
The scheme\cite{zpid} does not provide any false negative matches for single-compressed images in principle, however, it does not consider the errors caused by double-compression.
As a result, the scheme provided  some false negative matches for double-compressed images.
In the case of using the schemes \cite{ih1,itq}, hash values of similar images are close, so that they provided some false negative and positive matches under the simulation conditions.
Therefore, the proposed scheme is more effective than conventional ones in terms of false negative and positive matches, even if two images are very similar. 
\begin{table}[t!]
\caption{Querying performance.}
\label{tab:res}
\centering
\scalebox{1}{
\begin{tabular}{|@{\,}c@{\,}|@{\,}c@{\,}|c|c|c|c|c|c|c|}\hline
\multirow{1}{*}{scheme}&\multirow{1}{*}{database}&$p$[\%]&$r$[\%]\\\hline
\multirow{3}{*}{{proposed}}&$DB_1$&100&100\\\cline{2-4}
&$DB_2$&100&100\\\cline{2-4}
&$DB_3$&100&100\\\hline
\multirow{3}{*}{\shortstack{zero value\\ positions\cite{zpid}}}&$DB_1$&100&100\\\cline{2-4}
&$DB_2$&100&72.72\\\cline{2-4}
&$DB_3$&100&98.39\\\hline
\multirow{3}{*}{\shortstack{low-rank\\and sparse\\decomposition\cite{ih1}}}&$DB_1$&97.21&98.39\\\cline{2-4}
&$DB_2$&98.41&99.73\\\cline{2-4}
&$DB_3$&96.35&99.33\\\hline
\multirow{3}{*}{\shortstack{ITQ\cite{itq}}}&$DB_1$&67.24&99.33\\\cline{2-4}
&$DB_2$&67.67&99.87\\\cline{2-4}
&$DB_3$&62.98&98.79\\\hline
\end{tabular}}
\end{table}

\section{Conclusion}
It has been confirmed that the proposed identification scheme can identify robustly double-compressed images under the various conditions.
In this scheme, signs of only DC coefficients in DCT coefficients of Y component and one threshold value are used.
The use of them allows us to avoid the errors caused by double-compression.
The simulation results showed that the proposed scheme detected slightly differences and outperformed  other schemes including the state-of-art one, even if images were very similar.
\bibliography{ref}
\end{document}